------------------------------------------------------------------------ 
\documentstyle[eqsecnum,aps,epsf]{revtex}      

\begin{document}

\title{Dynamical calculation of direct muon-transfer rates from
thermalized muonic hydrogen  to C$^{6+}$ and O$^{8+}$}

\author{Renat A. Sultanov$^{1,2}$ and Sadhan K. Adhikari$^1$}

\address{$^1$Instituto de F\'\i sica Te\'orica,
Universidade Estadual Paulista,\\
01405-900 S\~{a}o Paulo, S\~{a}o Paulo, Brazil\\
$^2$Department of Physics, Texas A \& M University,\footnote{Present
address}
\\College Station, Texas
77843}

\date{\today}
\maketitle

\begin{abstract}

We perform dynamical calculations of direct muon-transfer rates from
thermalized muonic hydrogen isotopes to bare nuclei C$^{6+}$ and O$^{8+}$.
For these three-body charge-transfer reactions with Coulomb interaction in
the final state we use two-component integro-differential
Faddeev-Hahn-type equations in configuration space using close-coupling
approximation.  To take into account the  high polarizability of the
muonic hydrogen due to the large charge of the incident nuclei,
a  polarization potential is included in the
elastic channel. The large final-state Coulomb interaction is explicitly
included in the transfer channel. The transfer rates so calculated are in
good agreement with recent experiments.


\end{abstract}

\section{INTRODUCTION}
\label{sec:intro}

Mainly motivated by the possibility of muon-catalyzed fusion of the
hydrogen isotopes, theoretical and experimental
investigations of exotic
atomic and molecular systems involving negative muon ($\mu^-$) and
reactions in such systems continue to be  active fields of current
research \cite{np99}. Particular
attention is devoted to the study of the muon-transfer reaction from
muonic hydrogen  to other elements X$^{Z+}$ with large positive charge
$Z$, 
because such
reactions may
have large cross
sections and rates.  If the hydrogen is contaminated by even a small
amount of these heavier elements, this may strongly influence the process
of muon-catalyzed fusion by hydrogen isotopes. 
  Consequently, there has been considerable
recent experimental interest in the study of the muon-transfer reaction in
the collision of the muonic hydrogen by heavier nuclei, e.g. carbon
(C$^{6+}$), oxygen (O$^{8+}$), neon (Ne$^{10+}$), argon (Ar$^{18+}$)
\cite{pr97,jp93,hi98,pr95,pr98} etc. On the theoretical side, these  
three-body charge
transfer reactions involving a heavy transferred particle like muon and a
strong Coulomb interaction in the final state involving nuclei, such as
carbon and oxygen,  
continue to be 
challanging  problems calling for new investigations.   

Experimental study of muon-transfer rates from muonic hydrogen 
H$_\mu$   to   heavier nuclei, such as Ar$^{18+}$, Kr$^{36+}$
and Xe$^{54+}$, have revealed a smooth $Z$ dependence \cite{jetp66}.
In these
cases the transfer rate $\lambda_Z$  increases linearly  with $Z$.
Theoretical analyses are also in agreement with this conclusion
\cite{ann77}. However,   it has been found in  recent  experiment that the
predicted monotonic $Z$ dependence of the muon-transfer rate is not  valid
for all $Z$ \cite{pr95,hi93}. For these transfer rates, pronounced
fluctuations have been observed for elements up to argon contrary to the
smooth $Z$ dependence. The experimental
muon transfer rates for reactions like  
\begin{equation} (\mbox{H}
_\mu)_{1s} + {\mbox{X}^{Z+}} \rightarrow (\mbox{X}_\mu)^{(Z-1)+} +
\mbox{H}^+
\label{1}
\end{equation}  
depend in a complicated manner on the charge $Z$
\cite{hi93}.  Here $\mbox{H}$ stands for the hydrogen isotope $p$ (proton)
or $d$ (deuteron) and $\mbox{X}^{Z+}$ stands for the target nuclei.

Another phenomenon which has not yet found a satisfactory theoretical
explanation is the measured isotope effect, e.g.  the trend of the direct
muon-transfer  rate of reaction (\ref{1}) from $p\mu$ and
$d\mu$ to X$^{Z+}$, where X$^{Z+}$ represents C$^{6+}$, O$^{8+}$
\cite{jp93}, Ne$^{10+}$ \cite{hi93}, Ar$^{18+}$
\cite{pr97}, and Xe$^{54+}$ \cite{pr73}.  In cases of O$^{8+}$, Ar$^{18+}$
and Xe$^{54+}$ the direct-transfer rate decreases with increasing  mass
of the hydrogen isotope. Theoretical analyses \cite{ann77,ger63} also
support
this trend.  The experimental results for Ne$^{10}$ \cite{pr95,hi93} and
sulphur dioxide \cite{jp93} differ considerably from the theoretical
predictions. 

In view of this, here  we perform quantum dynamical calculation of
muon-transfer rates from  $p\mu$ to C$^{6+}$ and O$^{8+}$ and from 
$d\mu$ to  O$^{8+}$.
For this purpose, we use 
close-coupling approximation to two-component Faddeev-Hahn-type
dynamical equations \cite{hahn68,hahn72}.
We are currently investigating  muon-transfer rates
from muonic hydrogen  to other  heavier elements for a future
publication.

It is difficult to perform a quantum dynamical calculation of a
charge-transfer reaction.  In addition, a theoretical study of the
problems above is extremely complicated due to the large charge of these
heavy nuclei and the presence of a large number of open 
channels even at zero incident energy.  The large charge of the
nuclei
leads to  a  strong polarization of the muonic hydrogen in the
initial state and  a strong final-state Coulomb repulsion.  It is
difficult to incorporate these two effects properly in a dynamical
calculation. This is why there are {\it no}  dynamical calculations of
these muon-transfer rates.

Although there is a very large number of open channels in this problem,
for a given nuclei the muon is transferred predominantly to a few (muonic)
atomic labels of the heavy nuclei \cite{ann77,ger63}.  For example, muon
is captured mostly in the $n=4$ states of C$^{6+}$, and $n=5$ states of
O$^{8+}$.  Also these transfers take place mostly to the final
muonic-atomic states with low angular momenta and transfer rates are
negligible for atomic states with angular momenta $l>2$. The correct
dynamical formulation should include all the important transfer channels
and we included them in a previous study on muon transfer with light
nuclear targets \cite{pra,jpb,npa,epj99}.
However, it is quite impossible now to treat
even this reduced number of open transfer channels in a dynamical
calculation with heavier targets due to convergence difficulties in the
presence of the large polarization potential and large final-state Coulomb
interaction mentioned above. Hence in the present treatment we 
use a two-channel model  to 
calculate transfer to a
single
final state, where we include the elastic and one transfer channels.
Different sets of equations are used for the different final states.
Eventually, the total transfer rate is calculated by summing the different
contributions. 

The two-component Faddeev-Hahn-type equation, we use,
allows us to introduce explicitly a polarization potential in the initial
channel and the repulsive Coulomb potential in the final channel. This has
the advantage of building in the correct asymptotic behavior of the wave
function in a low-order close-coupling type approximation \cite{fb98}.
Hence as in
Ref. \cite{pra,jpb,npa} we make a two-state close-coupling approximation
to
the Faddeev-Hahn-type equation in the present study and find that a
numerical solution using the present scheme leads to very encouraging
agreement with experimental transfer rates.

In Sec. II  we present our formalism.
Numerical  results for muon-transfer rates  from  muonic hydrogen to
carbon
and oxygen are given in Sec. III and compared  with those of other
investigations. In Sec. IV we present a summary and outlook.

\section{Theoretical Formulation}

The theoretical framework for the present study will be based on the
formalism developed in Ref. \cite{jpb} which was used for the study of
muon transfer from muonic hydrogen atoms to light charged nuclei, such as,
He$^{2+}$ and Li$^{3+}$. Here we shall perform a similar study with
heavier charged nuclei, such as, C$^{6+}$ and O$^{8+}$. The presence of
the strong Coulomb interaction and the associated large polarization
interaction make the present calculational scheme far more complicated
theoretically and numerically compared to that of Ref. \cite{jpb}.  The
formalism of Ref. \cite{jpb} is a generalization over that of Ref.
\cite{pra} for charge-transfer reaction with no final-state Coulomb
interaction.  In the dynamical equations in
Ref. \cite{jpb} the final-state
Coulomb interaction is explicitly included in the transfer channel. In
addition, in the present work we explicitly include a 
polarization potential in the elastic channel.  In a coupled-channel
approach for atomic processes, the coupling to infinite number of p-wave  
states is
responsible for generating the polarization potential \cite{prs60}.
As it is impossible 
to include all such states in a numerical scheme, the commonly accepted
procedure is to replace these coupling terms by a
polarization potential as in Ref. \cite{rmp62}.
This idea has been recently used successfully in   
antiproton-hydrogen and antihydrogen-hydrogen reactions \cite{vor98}.
Next we describe 
the dynamical equations we
use based on the 
close-coupling approximation to    Faddeev-Hahn-type two-component
equations  \cite{pra}.

We use units 
 $e=\hbar=m_\mu=1$, where $m_\mu$
($e$) is
the muonic  mass  (charge), and
denote, the heavy nuclei  
(C$^{6+}$ or O$^{8+}$)
  by ${\sf 1}$, the hydrogen isotopes
by ${\sf 2}$ and muon by ${\sf 3}$.
Below the three-body breakup threshold, following 
two-cluster asymptotic configurations
are possible in the system {\sf 123}:  $({\sf 23})\ -\ {\sf 1}$ and
$({\sf 13})\ -\ {\sf 2}$. These two configurations 
  correspond to two distinct physical channels,  denoted by  
1 and 2, respectively. 
These configurations  
are  determined by the Jacobi coordinates
$(\vec r_{j3}, \vec \rho_k)$:
$\vec r_{13} = \vec r_3 - \vec r_1,
\hspace{6mm} \vec \rho_2 =
(\vec r_3 + m_1\vec r_1) / (1 + m_1) - \vec r_2$,
$\vec r_{23} = \vec r_3 - \vec r_2,
\hspace{6mm} \vec \rho_1 =
(\vec r_3 + m_2\vec r_2) / (1 + m_2) - \vec r_1$, where
$\vec r_{i}$, $m_{i}$ ($i=1, 2, 3,$) are coordinates and
masses of the particles respectively. 

Let us introduce
the total three-body wave function as a sum of two components
\begin{equation}
\Psi(\vec r_1, \vec r_2, \vec r_3) \ =\  \Psi_1 (\vec r_{23},\vec \rho_1)
\ + \ \Psi_2 (\vec r_{13},\vec \rho_2),
\label{eq:total2}
\end{equation}
where $\Psi_1 (\vec r_{23},\vec \rho_1)$
is quadratically integrable over the variable
$\vec r_{23}$, and  $\Psi_2 (\vec r_{13},\vec \rho_2)$ over the
variable $\vec r_{13}$. The components $\Psi_1$ and $\Psi_2$ carry the
asymptotic
boundary condition for channels 1 and 2, respectively.
The second component is responsible for
pure Coulomb interaction in the final state.
These components satisfy the following 
set of two coupled equations
\begin{eqnarray}\label{eq:1a}
[E\ - H_0 - V_{23}(\vec r_{23}) - U_{\scriptsize \mbox{pol}} 
(\vec \rho_1)]\Psi_1 (\vec
r_{23}, \vec \rho_1) &=&
[V_{23}(\vec r_{23})  + V_{12}(\vec r_{12}) - 
U_{\scriptsize \mbox C}\ (\vec \rho_2)]
\Psi_2 (\vec r_{13}, \vec \rho_2)\;,
\\
\vspace{2mm}
[E - H_0 - V_{13}(\vec r_{13})-\ U_{\scriptsize \mbox C}\ (\vec \rho_2)]
\Psi_2 (\vec r_{13}, \vec \rho_2) &=&
[V_{13} (\vec r_{13})+ V_{12}(\vec r_{12})
-U_{\scriptsize \mbox{pol}}(\vec
\rho_1)]\Psi_1 (\vec r_{23}, \vec \rho_1)\;,
\label{eq:1}
\end{eqnarray}
where $E$ is the center-of-mass energy, $H_0$ is the total kinetic energy
operator,  $V_{ij} (\vec r_{ij})$
pair potentials $(i \not= j = 1, 2, 3)$, 
$U_{\scriptsize \mbox C}$  is the final-state Coulomb  interaction given
by
\begin{equation}
U_{\scriptsize \mbox C}(\vec \rho_2) = \frac{(Z - 1)Z'}{\rho_2}\ \ ,
\end{equation}
with $Z$ the charge of the heavy nuclei  and $Z'(=1)$  the
charge of the hydrogen  isotope. 
Here $U_{\scriptsize \mbox{pol}}$ is the
 polarization potential given by 
 \cite{ger63}
\begin{equation}
U_{\scriptsize \mbox{pol}}(\vec \rho_1) = -\frac{9Z^2}{4\rho_1^4}\ \ ,
\hskip 0.5cm \rho_1
>
\Lambda
\label{eq:pol}
\end{equation} 
and zero otherwise. The value of the cut-off parameter $\Lambda$ has to be
chosen appropriately (see Sec. III).
By adding Eqs.  (\ref{eq:1a}) and (\ref{eq:1}) we find that they are
equivalent to the Schr\"odinger equation.

Distortion potential  has been very useful in model
and phenomenological description of reaction and scattering in nuclear
\cite{t61} and
atomic physics \cite{mott65}. Although such distortion potentials are
unnecessary in a
complete solution of the Schr\"odinger equation, they enhance the
agreement with experiment in a simplified model description. For example,
a long-range polarization (distortion) potential has been routinely used
in electron-atom scattering \cite{rmp62,nes}. Such a distortion  potential
arising from 
the polarization of the muonic hydrogen isotope due to the bare nuclei
is effective in
the initial channel and has been included in Eq. (\ref{eq:1a}). This
polarization potential for C$^{6+}$ or O$^{8+}$ is much stronger and its
effect on cross sections much more pronounced than in the case of electron
scattering. Hence, for obtaining a better agreement with experiment  in a
model calculation it is prudent to include the polarization potential in
the elastic channel.
 There
should also be such a polarization potential in the final channel.
However, by far  the
most important interaction in the final channel  is the Coulomb potential
between the proton (or deuteron) and the charged muonic atom
$(X_\mu)^{(Z-1)+}$. 
This Coulomb (distortion) potential has also been explicitly included in
Eq. (\ref{eq:1}). This will help in obtaining a realistic description of
the transfer process as we shall find in the following.

Because of the strong final-state Coulomb interaction in the present
muon-transfer problems it is 
very difficult to develop and solve successfully multichannel
models 
based on Eqs. (\ref{eq:1a}) and  (\ref{eq:1}) above as in Ref. \cite{jpb}.
Hence, 
for solving Eqs. (\ref{eq:1a}) and  (\ref{eq:1}) we expand the wave
function
components in terms of bound states in initial and final channels,
and project these equations on these bound states. The expansion of the
wave
function is given by
\begin{equation}
\Psi_1(\vec r_{23}, \vec \rho_{1}) \approx
\frac{f_{1s}^{(1)}(\rho_1)}{\rho_1}
R_{1s,\mu_1}^{(Z')}(|\vec r_{23}|)/4\pi,
\label{eq:expan1}
\end{equation}
\begin{equation}
\Psi_2(\vec r_{13}, \vec \rho_{2}) \approx
\frac{f_{nl{\cal L}}^{(2)}(\rho_2)}{\rho_2}
R_{nl,\mu_2}^{(Z)}(|\vec r_{13}|)
\left \{ Y_{\cal L}(\hat \rho_2) \otimes
Y_l(\hat r_{13}) \right \}_{00},
\label{eq:expan2}
\end{equation}
where $nl{\cal L}$ are quantum numbers of the
three-body final-state,
$\mu_1=m_3m_2/(m_3+m_2)$, $\mu_2=m_3m_1/(m_3+m_1)$,
$Y_{lm}$'s are
the  spherical harmonics,
$R_{nl,\mu_i}^{(Z)}(|\vec r|)$
is the radial part of the
hydrogen-like bound-state wave function for reduced mass $\mu_i$
and charge $Z$,
$f_{1s}^{(1)}(\rho_1)$ and
$f_{nl{\cal L}}^{(2)}(\rho_2)$ are the unknown expansion coefficients.
This prescription is similar to that adopted in the close-coupling
approximation. After a proper angular momentum projection,
the  set of two-coupled integro-differential equations
for the unknown expansion functions can be written as 
\begin{eqnarray}
\left[ (k_1^{(1)})^2\ +\ \frac{\partial^2}
{\partial \rho_1^2}\ - \
2M_1U_{\scriptsize \mbox{pol}} (\vec \rho_1) \right]
f_{1s}^{(1)}(\rho_1) =
g_1\sqrt{(2{\cal L} + 1)}
\int_{0}^{\infty} d \rho_{2}
f_{nl{\cal L}}^{(2)}(\rho_{2})
\; \nonumber \\
\int_{0}^{\pi}d \omega \sin\omega
R_{1s,\mu_1}^{(Z')}(|\vec{r}_{23}|)
\left(-\frac{Z'}{|\vec {r}_{23}|} + \frac{Z}{|\vec {r}_{12}|}
 - U_{\scriptsize \mbox C} (\vec \rho_2) \right)
R_{nl,\mu_2}^{(Z)}(|\vec{r}_{13}|)
\;\nonumber \\
\rho_1 \rho_2
C_{{\cal L} 0l0}^{00}
Y_{lm}(\nu_{2}, \pi)/\sqrt{4\pi}\;,
\label{eq:most1}
\end{eqnarray}
\begin{eqnarray}
\left[ (k^{(2)}_n)^2\ +\ \frac{\partial^2}
{\partial \rho_2^2}\ -\
\frac{{\cal L} ({\cal L} + 1)}{\rho_2^2} -
2M_2U_{\scriptsize \mbox C} (\vec \rho_2)
\right] f_{nl{\cal L}}^{(2)}(\rho_2) = g_2
\sqrt{(2{\cal L }+ 1)}
\; \nonumber \\
\int_{0}^{\infty} d \rho_{1}
f_{1s}^{(1)}(\rho_{1})\int_{0}^{\pi}
d \omega \sin\omega
R_{nl,\mu_2}^{(Z)}(|\vec{r}_{13}|)
\left(-\frac{Z}{|\vec {r}_{13}|} + \frac{Z}{|\vec {r}_{12}|}
- U_{\scriptsize \mbox{pol}} (\vec \rho_1)\right)
\;\nonumber \\
R_{1s,\mu_1}^{(Z')}(|\vec{r}_{23}|)
\rho_{2} \rho_1
C_{{\cal L }0l0}^{00}
Y_{lm}(\nu_{1}, \pi)/\sqrt{4\pi}\;.
\label{eq:most2}
\end{eqnarray}
Here 
$k_1^{(1)}   = \sqrt{2M_1{(E- E_{1s}^{(2)})}}$,
$k_n^{(2)} = \sqrt{2M_2{(E- E_n^{(1)})}}$
with
$M_1^{-1}= m_1^{-1} + (1 + m_2)^{-1}$ and
$M_2^{-1}= m_2^{-1} + (1 + m_1)^{-1}$,
$E_n^{(j)}$ is the binding energy of pair $(j3)$ and
$g_j=4\pi M_j/\gamma^{3}$ ($j = 1, 2$), 
$\gamma=1-m_1m_2/((1+m_1)(1+m_2))$,
$C_{{\cal L} 0lm}^{Lm}$ the Clebsch-Gordon coefficient,
$\omega$ is the angle between the Jacobi coordinates
$\vec \rho_1$ and $\vec \rho_2$, $\nu_1$ is the angle between 
$\vec r_{23}$ and $\vec \rho_1$, $\nu_{2}$ is the angle
between $\vec r_{13}$ and $\vec \rho_{2}$.

We find that after the projection of the Faddeev-Hahn-type  
equations  (\ref{eq:1a}) and (\ref{eq:1}) on the basis states, the
initial-state polarization and the final-state Coulomb potentials survive
on the left-hand-side of the resultant equations (\ref{eq:most1}) and
(\ref{eq:most2}). The
presence of the explicit Coulomb potential in the final channel will
automatically yield the correct physical Coulomb-wave boundary  condition
in that 
channel. The explicit inclusion of the polarization potential,
although  has no
effect on the boundary condition in the initial channel, substantially
improves the the results of the truncated model calculation based on 
Eqs. (\ref{eq:most1}) and (\ref{eq:most2}) as we shall see in the next
section.

To find unique solution to Eqs. 
(\ref{eq:most1})$-$(\ref{eq:most2}),
appropriate boundary conditions are to be considered.  We impose the
usual condition of regularity at the origin 
$f_{1s}^{(1)}(0)\mathop{\mbox{\large$=$}}0$ and
$f_{nl\cal L}^{(2)}(0)\mathop{\mbox{\large$=$}}0$.
Also
for the present scattering  problem with $1 +(23)$ as the initial state,
in the asymptotic region, two solutions to Eqs.
(\ref{eq:most1})$-$(\ref{eq:most2}) satisfy the following boundary
conditions
\begin{eqnarray}
f_{1s}^{(1)}(\rho_1)
&\mathop{\mbox{\large$\sim$}}\limits_{\rho_1 \rightarrow + \infty}&
\sin(k^{(1)}_1\rho_1) + {\it K}_{11}^{nl}\cos(k^{(1)}_1\rho_1)\;,
\\
f_{nl\cal L}^{(2)}(\rho_2)
&\mathop{\mbox{\large$\sim$}}\limits_{\rho_2 \rightarrow + \infty}&
\sqrt{v_1 / v_2}{\it K}_{12}^{nl}
\cos(k^{(2)}_1\rho_2 - \eta / 2k^{(2)}_1 \ln2k^{(2)}_1\rho_2
-\pi{\cal L}/2)\;,
\label{eq:cond88}
\end{eqnarray}
where 
$\it K_{ij}^{nl}$ are the appropriate coefficients.
For scattering
with ${\sf 2} + ({\sf 13})$ as the initial state, we have the following
conditions
\begin{eqnarray}
f_{1s}^{(1)}(\rho_1)
\mathop{\mbox{\large$\sim$}}\limits_{\rho_1 \rightarrow + \infty}
\sqrt{v_2 / v_1}{\it K}_{21}^{nl}\cos(k^{(1)}_1\rho_1)\;,
\\
f_{nl\cal L}^{(2)}(\rho_2)
\mathop{\mbox{\large$\sim$}}\limits_{\rho_2 \rightarrow + \infty}
\sin(k^{(2)}_1\rho_2 - \eta / 2k^{(2)}_1 \ln2k^{(2)}_1\rho_2
-\pi{\cal L}/2)+\\
{\it K}_{22}^{nl} \cos(k^{(2)}_1\rho_2 -
\eta / 2k^{(2)}_1 \ln2k^{(2)}_1\rho_2 - \pi{\cal L}/2)\;,
\label{eq:cond8888}
\end{eqnarray}
where $v_i$ ($i=1,2$) is  velocity in channel $i$. 
The Coulomb parameter in the second transfer channel
is $\eta = 2M_2(Z-1)/k^{(2)}_n$  \cite{mott65}.
The coefficients $\it K_{ij}^{nl}$ are obtained from the
numerical solution of the Faddeev-Hahn-type equations.
The cross sections are given by
\begin{eqnarray}
\sigma^{\scriptsize \mbox{tr}}_{1s\rightarrow nl}\ =\
\frac{4\pi}{k^{(1)2}}\frac{({\it K}_{12}^{nl})^2}
{(D - 1)^2 + ({\it K}_{11}^{nl} + {\it K}_{22}^{nl})^2},
\end{eqnarray}
where $D = {\it K}_{11}^{nl}{\it K}_{22}^{nl}
-{\it K}_{12}^{nl}{\it K}_{21}^{nl}$. When $k^{(1)} \rightarrow 0$:
$\sigma^{\scriptsize \mbox{tr}}_{1s\rightarrow nl} \sim 1/k_1^{(1)}$.
The transfer rates  are defined by 
\begin{equation}
\lambda^{\scriptsize \mbox{tr}}_{1s\rightarrow nl}\ 
=\ \sigma^{\scriptsize \mbox{tr}}_{1s\rightarrow nl} v_1 N_0,
\label{eq:tr1}
\end{equation}
with $v_1$ being the relative velocity of the incident
fragments and 
$N_0$ the liquid-hydrogen density chosen here
as $4.25\times10^{22}$ $\mbox{cm}^{-3}$, because
$\lambda^{\scriptsize \mbox{tr}}(k^{(1)}\rightarrow 0) \sim $ const.
In our model approach the total muon transfer rate is
\begin{equation}
\lambda_{\scriptsize \mbox{tot}}^{\scriptsize \mbox{tr}} = 
\sum \lambda^{\scriptsize \mbox{tr}}_{1s\rightarrow nl}.
\label{eq:tr2}
\end{equation}

\section{Numerical Results}

We employ muonic atomic unit: distances are measured in units of $a_\mu$,
where $a_\mu$ is the radius of muonic hydrogen atom.  The
integro-differential equations are solved by discretizing them into a
linear system of equations. The integrals in Eqs. (\ref{eq:most1})  and
(\ref{eq:most2}) are discretized using the trapezoidal rule and the
partial derivatives are discretized using a three-point rule \cite{as}.
The discretized equation is subsequently solved by Gauss elimination
method.  As we are concerned with the low-energy limit, only the total
angular momentum $L=0$ is taken into account.  Even at zero incident
energy, the transfer channels are open and their wave functions are
rapidly oscillating Coulomb waves. In order to get a converged solution we
needed a large number of discretization points. More points are taken near the
origin where the interaction potentials are large; a smaller number of
points are needed at large distances.

First we solved the system of equations without the polarization potential
in the incident channel. However, the final-state Coulomb interaction  is
correctly represented in our model. In this case it is relatively
easy to obtain
the numerical convergence for the system of equations which includes the
elastic channel and one transfer channel at a time. Finally, the total
transfer cross section is calculated by adding the results of different
two-channel contributions. 
Without the polarization potential, we needed up
to 700 discretization points adequately distributed between 0 and
50$a_\mu$. Near the origin we took up to 60 equally spaced points per unit
interval ($a_\mu$). 

It was more difficult to obtain convergence with the polarization
potential. 
The polarization potential (\ref{eq:pol}) is taken to be zero at small
distances below the cut off $\Lambda$. In this case to get numerical
convergence we had to integrate to very large distances $-$ up to
300$a_{\mu }$. We needed up to 2000 discretization points to obtain
convergence. Again we needed  more points near the origin and
less at large distances.  For example, near the origin we took up to 60
equally spaced points per  unit length interval $a_\mu$; in the
intermediate region ($\rho = 10 - 20 a_\mu$) we took up to 15 equally
spaced points per unit length interval, and in the asymptotic
region ($\rho = 20 - 300 a_\mu$) we took up to 5 equally spaced points
per unit length interval.  It is well-known that the results for
the cross section is sensitive to the value of the cut off $\Lambda$ of
the polarization potential. The short-range potential of the present
problem extends to about 25$a_\mu$. We considered the polarization
potential in the asymptotic region $\rho_1 > \Lambda \simeq 75a_\mu$. For
a small variation of $\Lambda$ in this region from $75a_\mu$ to about
120$a_\mu$, we find the transfer cross sections to be reasonably constant
and the reported transfer cross sections of this study are the averages of
these cross sections. 
If $\Lambda$ is increased past $120a_\mu$, the effect of the
polarization potential on the cross sections gradually decreases and
finally disappears. If $\Lambda$ is decreased much below $75a_\mu$, the
cross sections become rapidly varying function of $\Lambda$ and could
become unphysically large. The range of $\Lambda $ values (here between
$75a_\mu$ and $120a_\mu$) for which the cross sections are  slowly
varying smooth functions should increase  with the charge of the bare
nucleus.

We present partial muon-transfer rates 
$\lambda^{\scriptsize \mbox{tr}}_{nl}$  and
total transfer rates 
$\lambda_{\scriptsize \mbox{tot}}^{\scriptsize \mbox{tr}}$ 
calculated using the
formulation of last section.  
In this work using the model of Sec. II we calculate the low-energy
muon-transfer rates from ($p\mu$)$_{1s}$ to C$^{6+}$ and O$^{8+}$
and from  ($d\mu$)$_{1s}$ to  O$^{8+}$.
From other theoretical \cite{ann77,ger63} investigations it
was concluded that in the case of C$^{6+}$ the transfer takes place
predominantly to the $n=4$ state and for O$^{8+}$ it happens to the $n=5$
state, which is also found to be true in our model calculation. Hence in
this work we
only present rates for the $l=0,1,2$ states of the $n=4$ and 5 orbitals of
carbon and oxygen, respectively. The contribution of the higher angular
momentum states to the total transfer cross section is very small. 
Numerically converged results were obtained in these cases. The low energy
partial rates
$\lambda^{\scriptsize \mbox{tr}}_{1s\rightarrow nl}$ $/10^{10}$
sec$^{-1}$ and
total rates
$\lambda_{\scriptsize \mbox{tot}}^{\scriptsize \mbox{tr}}$ $/10^{10}$
sec$^{-1}$ 
are presented in
Tables I,
II, and III together with the results of other theoretical and
experimental works.

First we comment on the results in Table I for muon transfer from 
($p\mu$)$_{1s}$  to C$^{6+}$.  The partial
transfer rates
without the polarization potential increases with decreasing
center-of-mass energy $E$ and saturates to a constant value for $E<0.04$
eV for 4s, 4p, and 4d states of muonic carbon. In the case of C$^{6+}$ the
transfer takes place predominantly to the 4s state. These qualitative
behaviors are also true after the inclusion of the polarization potential
and were also true in a previous theoretical study. However, after the
inclusion of the polarization potential the 4s and 4p
transfer rates are enhanced by about a factor of 1.5.
The
present total transfer rate of $8.5\times 10^{10}$ sec$^{-1}$ is about
three times larger than the previous theoretical calculation of 
 $2.8\times 10^{10}$ sec$^{-1}$ \cite{ger63}.  
We quote two experimental results in this case: $(9.5\pm 0.5)\times
10^{10}$ sec$^{-1}$ \cite{pr97} and $(5.1\pm 1.0)\times
10^{10}$ sec$^{-1}$ \cite{jetp66}. The present theoretical result lies in
between these two somewhat conflicting experimental results.

In the case of muon transfer from ($p\mu$)$_{1s}$ to O$^{8+}$, we
find from Table II that transfer takes place predominantly to the 5s and 
5p
states of muonic oxygen. Again the transfer rates saturate and
attain constant values for $E<0.04$ eV.  The transfer
rate is higher in the 5s state and lowest in the 5d state. This behavior
remain true after the inclusion of the
polarization potential, when
the transfer rate to the 5p state increases by a factor of more than two
whereas the contribution to the 5s state increases by a factor of 1.5.
The
present total transfer rate of $(7.7\pm 0.5)\times 10^{10}$ sec$^{-1}$ is
about 1.5 times larger than the previous theoretical calculation of
$5.6\times 10^{10}$ sec$^{-1}$ \cite{ger63}
and in reasonable agreement with the recent experimental rate of $(8.5\pm
0.2)\times 10^{10}$ sec$^{-1}$ \cite{hi98}.

Finally, in the case of muon transfer from ($d\mu$)$_{1s}$
to O$^{8+}$, we find from Table III that transfer also takes place
predominantly to the 5s state of muonic oxygen. The contribution to
the transfer rate due to 5s state is two times as large as the
contribution due to the 5p state.
Again the transfer rate saturate and attain
a constant value for $E<0.04$ eV.
  After the inclusion of the polarization
potential the transfer rate to the 5s state increases by a factor of 1.5.
The present total transfer rate of $(4.4\pm 0.6)\times 10^{10}$
sec$^{-1}$ is in reasonable agreement with 
the recent experimental rate of $5.5\times 10^{10}$ sec$^{-1}$
\cite{jp93}.

The $Z$ dependence of the transfer rates from a specific hydrogen isotope
to X$^{Z+}$ has been a subject of interest. Although for large $Z$ these
rates increase linearly with $Z$, there is no general behavior for small
$Z$. The most recent experimental transfer rates decrease when we move
from the system
$p\mu$ $-$C$^{6+}$ to  $p\mu$ $-$O$^{8+}$ \cite{pr97,hi98}. 
Through our dynamical
calculation
we have been able to reproduce this behavior. Our calculation is also
consistent with the experimentally observed isotope effect, e.g., 
the transfer rate decreases when we move from
$p\mu$ $-$O$^{8+}$ to  $d\mu$ $-$O$^{8+}$ \cite{jp93,hi98}.

\section{Conclusion}

We have studied muon-transfer reactions from muonic hydrogen to carbon and
oxygen nuclei employing a full quantum-mechanical few-body description of
rearrangement scattering by solving the Faddeev-Hahn-type equations using
close-coupling approximation.  To provide the correct asymptotic form of
the wave function in the transfer channel, the final-state Coulomb
interaction has been incorporated directly into the equations. We also
included a  polarization potential at large distances in
the initial channel. 
It is shown that in the present approach, the
application of a close-coupling-type ansatz leads to satisfactory results
for direct muon-transfer reactions
from muonic hydrogen to ${\mbox {C}}^{6+}$ and ${\mbox {O}}^{8+}$.  In the
case of muon transfer from ($p\mu$)$_{1s}$ to C$^{6+}$ the present
transfer rate to the 4s state of muonic carbon is about 1.5 times larger
than that to the 4p state.
For muon transfer from ($p\mu$)$_{1s}$ to
O$^{8+}$, the present transfer rate to the 5s state of oxygen is about
twice as large as that to the 5p state.
The present rates are much larger by factors of about
two to three compared to  the calculation of Ref. \cite{ger63}.
Finally, in the
case of muon transfer from ($d\mu$)$_{1s}$ to O$^{8+}$, the present
transfer rate to the 5s state is  large compared to that to the 5p
state.
In all cases the inclusion of the polarization potential improves the
agreement with experiment and our final transfer rates 
 are in encouraging agreement with
recent experiments \cite{pr97,jp93,hi98}.
The present rates for oxygen from ($p\mu$)$_{1s}$ and 
($d\mu$)$_{1s}$ are in agreement with the observed isotope effect
\cite{jp93}:  the transfer rate increases with the decrease of the mass of
the hydrogen isotope. 
Because of the present promising results for the muon-transfer rates for
$Z=6$ and $Z=8$ it seems useful to make future applications of the present
formulation for larger targets. Calculations involving nuclei of higher
charges (Ne$^{10+}$, S$^{16+}$, Ar$^{18+}$ etc.) are in progress. 

\acknowledgments
We acknowledge the support from Funda\c{c}\~{a}o
de Amparo \~{a} Pesquisa do Estado de S\~{a}o Paulo of  Brazil.
The numerical calculations were performed on the IBM SP2
Supercomputer of the Departamento de
F\'\i sica - IBILCE - UNESP,
S\~{a}o Jos\'e do Rio Preto, Brazil.

\mediumtext

\begin{table}
{Table I. Low energy partial
$\lambda^{\scriptsize \mbox{tr}}_{1s\rightarrow
nl}$$/10^{10}$$\mbox{sec}^{-1}$
and total 
$\lambda_{\scriptsize
\mbox{tot}}^{\scriptsize \mbox{tr}}/10^{10}$$\mbox{sec}^{-1}$
muon transfer rates reduced to liquid-hydrogen density
$N_0 = 4.25\times10^{22}$ $\mbox{cm}^{-3}$ from muonic
hydrogen ($p\mu$)$_{1s}$
to hydrogen-like excited state of muonic carbon
(${{\mbox {C}}}_\mu$)$^{5+}_{n=4}$.}
\begin{tabular}{lcccccccclccccccc}
\multicolumn{1}{l}{Energy}                &
\multicolumn{1}{c}{ }                     &
\multicolumn{2}{c}{$U_{\scriptsize \mbox{pol}}(\rho_1) = 0$} &
\multicolumn{2}{c}{With polarization} &
\multicolumn{1}{c}{Theory}                &
\multicolumn{2}{c}{Experiment}\\
\multicolumn{1}{l}{$E$ (eV)}              &
\multicolumn{1}{l}{$(nl)$}                &
\multicolumn{1}{c}{$\lambda^{\scriptsize \mbox{tr}}_{1s\rightarrow nl}$} &
\multicolumn{1}{c}{$\lambda_{\scriptsize
\mbox{tot}}^{\scriptsize \mbox{tr}}$} &
\multicolumn{1}{c}{$\lambda^{\scriptsize \mbox{tr}}_{1s\rightarrow nl}$} &
\multicolumn{1}{c}{$\lambda_{\scriptsize
\mbox{tot}}^{\scriptsize \mbox{tr}}$} &
\multicolumn{1}{c}{\hspace{2mm}{\cite{ger63}}}               &
\multicolumn{1}{c}{\hspace{2mm}{\cite{jetp66}}}                &
\multicolumn{1}{c}{\hspace{2mm}{\cite{pr97}}}              &
\\ \hline
$ $
& $ 4s $ & $ 3.4 $ & $ $ & $ 5.2 \pm 0.2 $
& $ $ & $ $ & $ $\\
$  0.04 $
& $ 4p $ & $ 2.1 $ & $ 5.5 $ & $ 3.2 \pm 0.5 $
& $ 8.5 \pm 0.7 $ & $ 2.8 $ & $ 5.1 \pm 1.0 $ & $ 9.5 \pm 0.5 $\\
$ $
& $ 4d $ & $ 0.05 $ & $ $ & $ 0.1 $
& $ $ & $ $ & $ $\\
\hline
$ $
& $ 4s $ & $ 2.1 $ & $ $ & $ 2.8 \pm 0.2$
& $ $ & $ $ & $ $\\
$ 0.1 $
& $ 4p $ & $ 1.1 $ & $ 3.2 $ & $ 1.5 \pm 0.2$
& $ 4.3 \pm 0.4 $ & $  $ & $  $ & $  $\\
$ $
& $ 4d $ & $ \sim 0 $ & $ $ & $ \sim 0 $
& $ $ & $ $ & $ $\\
\hline
$ $
& $ 4s $ & $ 1.2 $ & $ $ & $ 1.6 \pm 0.1$
& $ $ & $ $ & $ $\\
$ 0.5 $
& $ 4p $ & $ 0.4 $ & $ 1.6 $ & $ 0.7 \pm 0.1$
& $ 2.3 \pm 0.2$ & $  $ & $  $ & $  $\\
$ $
& $ 4d $ & $ \sim 0 $ & $ $ & $ \sim 0 $
& $ $ & $ $ & $ $\\
\end{tabular}
\end{table}

\mediumtext

\begin{table}
{Table II. Low energy partial
$\lambda^{\scriptsize \mbox{tr}}_{1s\rightarrow
nl}$$/10^{10}$$\mbox{sec}^{-1}$
and total 
$\lambda_{\scriptsize
\mbox{tot}}^{\scriptsize \mbox{tr}}/10^{10}$$\mbox{sec}^{-1}$
muon transfer rates reduced to liquid-hydrogen density
$N_0 = 4.25\times10^{22}$ $\mbox{cm}^{-3}$ from muonic
hydrogen ($p\mu$)$_{1s}$
to hydrogen-like excited state of muonic oxygen
(${{\mbox {O}}}_\mu$)$^{7+}_{n=5}$.}
\begin{tabular}{lcccccccclccccccc}
\multicolumn{1}{l}{Energy}                &
\multicolumn{1}{c}{ }                     &
\multicolumn{2}{c}{$U_{\scriptsize \mbox{pol}}(\rho_1) = 0$} &
\multicolumn{2}{c}{With polarization} &
\multicolumn{1}{c}{Theory}                &
\multicolumn{2}{c}{Experiment}\\
\multicolumn{1}{l}{$E$ (eV)}              &
\multicolumn{1}{l}{$(nl)$}                &
\multicolumn{1}{c}{$\lambda^{\scriptsize \mbox{tr}}_{1s\rightarrow nl}$} &
\multicolumn{1}{c}{$\lambda_{\scriptsize
\mbox{tot}}^{\scriptsize \mbox{tr}}$} &
\multicolumn{1}{c}{$\lambda^{\scriptsize \mbox{tr}}_{1s\rightarrow nl}$} &
\multicolumn{1}{c}{$\lambda_{\scriptsize
\mbox{tot}}^{\scriptsize \mbox{tr}}$} &
\multicolumn{1}{c}{\hspace{2mm}{\cite{ger63}}}               &
\multicolumn{1}{c}{\hspace{2mm}{\cite{jp93}}}                &
\multicolumn{1}{c}{\hspace{2mm}{\cite{hi98}}}                &
\\ \hline
$ $
& $ 5s $ & $ 3.5 $ & $ $ & $ 5.5 \pm 0.2 $
& $  $ & $ $ & $ $\\
$ 0.04 $
& $ 5p $ & $ 0.8 $ & $ 4.35 $ & $ 2.1 \pm 0.2 $
& $ 7.7 \pm 0.5 $ & $ 5.6 $ & $ 8.3 $ & $ 8.5 \pm 0.2 $\\
$ $
& $ 5d $ & $ 0.05 $ & $ $ & $ 0.1 \pm 0.05 $
& $ $ & $ $ & $ $\\
\hline
$ $
& $ 5s $ & $ 3.1 $ & $ $ & $ 5.0 \pm 0.2$
& $ $ & $ $ & $ $\\
$ 0.1 $
& $ 5p $ & $ 0.7 $ & $ 3.8 $ & $ 1.7 \pm 0.2 $
& $ 6.8 \pm 0.5 $ & $  $ & $  $\\
$ $
& $ 5d $ & $ 0.02 $ & $ $ & $ 0.05  $
& $ $ & $ $ & $ $\\
\hline
 $ $
 & $ 5s $ & $ 2.0 $ & $ $ & $ 2.9 \pm 0.1$
 & $ $ & $ $ & $ $\\
 $ 0.5 $
 & $ 5p $ & $ 0.2 $ & $ 2.2 $ & $ 1.0 \pm 0.1 $
 & $ 3.9 \pm 0.2 $ & $  $ & $  $\\
 $ $
 & $ 5d $ & $ \sim 0 $ & $ $ & $ \sim 0 $
 & $ $ & $ $ & $ $
\end{tabular}
\end{table}
\mediumtext

\begin{table}
{Table III. Low energy partial
$\lambda^{\scriptsize \mbox{tr}}_{1s\rightarrow
nl}$$/10^{10}$$\mbox{sec}^{-1}$
and total 
$\lambda_{\scriptsize
\mbox{tot}}^{\scriptsize \mbox{tr}}/10^{10}$$\mbox{sec}^{-1}$
muon transfer rates reduced to liquid-hydrogen density
$N_0 = 4.25\times10^{22}$ $\mbox{cm}^{-3}$
from muonic hydrogen ($d\mu$)$_{1s}$
to hydrogen-like excited state of muonic oxygen
(${{\mbox {O}}}_\mu$)$^{7+}_{n=5}$.}
\begin{tabular}{lcccccccclccccccc}
\multicolumn{1}{l}{Energy}                &
\multicolumn{1}{c}{ }                     &
\multicolumn{2}{c}{$U_{\scriptsize \mbox{pol}}(\rho_1) = 0$} &
\multicolumn{2}{c}{With polarization} &
\multicolumn{1}{c}{Experiment}\\
\multicolumn{1}{l}{$E$ (eV)}              &
\multicolumn{1}{l}{$(nl)$}                &
\multicolumn{1}{c}{$\lambda^{\scriptsize \mbox{tr}}_{1s\rightarrow nl}$} &
\multicolumn{1}{c}{$\lambda_{\scriptsize
\mbox{tot}}^{\scriptsize \mbox{tr}}$} &
\multicolumn{1}{c}{$\lambda^{\scriptsize \mbox{tr}}_{1s\rightarrow nl}$} &
\multicolumn{1}{c}{$\lambda_{\scriptsize
\mbox{tot}}^{\scriptsize \mbox{tr}}$} &
\multicolumn{1}{c}{\hspace{2mm}{\cite{jp93}}}                &
\\ \hline
$ $
& $ 5s $ & $ 1.9 $ & $ $ & $ 2.9 \pm 0.3 $
& $  $ & $ $ & $ $\\
$ 0.04 $
& $ 5p $ & $ 0.8 $ & $ 2.7 $ & $ 1.5 \pm 0.3 $
& $ 4.4 \pm 0.6 $ & $ 5.5 $\\
$ $
& $ 5d $ & $ \le 0.01 $ & $ $ & $ \sim 0 $
& $ $ & $ $ & $ $\\
\hline
$ $
& $ 5s $ & $ 1.1 $ & $ $ & $ 1.8 \pm 0.2 $
& $ $ & $ $ & $ $\\
$ 0.1 $
& $ 5p $ & $ 0.5 $ & $ 1.6 $ & $ 0.7 \pm 0.2 $
& $ 2.5 \pm 0.4 $ & $  $\\
$ $
& $ 5d $ & $ \le 0.01 $ & $ $ & $ \sim 0 $
& $ $ & $ $ & $ $\\
\hline
 $ $
 & $ 5s $ & $ 0.7 $ & $ $ & $ 1.0 \pm 0.2$
 & $ $ & $ $ & $ $\\
 $ 0.5 $
 & $ 5p $ & $  0.1 $ & $ 0.8 $ & $ 0.2 \pm 0.1 $
 & $ 1.2 \pm 0.3 $ & $  $\\
 $ $
 & $ 5d $ & $ \le 0.01 $ & $ $ & $ \sim 0 $
 & $ $ & $ $ & $ $
\end{tabular}
\end{table}
\end{document}